\documentclass[aps,prl,twocolumn,superscriptaddress,showpacs]{revtex4}
\usepackage{graphicx}

\input epsf.sty
\begin{document}
\title{Pairing Symmetry and Upward Curvature of Upper Critical Field
in Superconducting Na$_{0.35}$CoO$_{2} \cdot y$H$_{2}$O}
\author{Jun-Ting Kao}
\affiliation{ Department of Physics, National Tsing Hua University,
Hsinchu 30043, Taiwan}
\author{Jiunn-Yuan Lin}
\affiliation{ Institue of Physics, National Chao Tung University,
Hsinchu 30043, Taiwan}
\author{Chung-Yu Mou}
\affiliation{
 Department of Physics, National Tsing Hua University,
Hsinchu 30043, Taiwan} \affiliation{Physics Division, National
Center for Theoretical Sciences, P.O.Box 2-131, Hsinchu, Taiwan}
\date{\today}
\begin{abstract}
The origin for the upward curvature of the upper critical field ($H_{c2}$)
observed in hydrate cobaltate Na$_{0.35}$CoO$%
_{2} \cdot y$H$_{2}$O is investigated based on the microscopic gap
equation. It is shown that the observed upward curvature results
from the transition between two different pairing symmetries that
occur on different energy bands. Furthermore, different pairing
symmetries involved in the transition results in different upward
curvatures. By considering transitions among all lowest possible
pairing symmetries, it is found that the transition of the pairing
symmetry from $s$-wave at low fields to $d_{x^2-y^2}+ id_{xy}$ at
high fields is the best fit to the experimental data. Our results
provide an important clue to the understanding of the
superconductivity in hydrate cobaltate.
\end{abstract}
\pacs{73.20.-r, 73.20.At, 73.21.Hb} \maketitle

Since the discovery of superconductivity in hydrate cobaltate Na$_{0.35}$CoO$%
_{2} \cdot y$H$_{2}$O\cite{Takada}, extensive theoretical and
experimental studies have been devoted to elucidate the mechanism of
superconductivity. To unravel the mechanism, identifying the
underlying pairing symmetry would be the first step. The
two-dimensional triangle lattice formed by CoO$_2$ provides an
alternative lattice symmetry to the square CuO$_2$ lattice in high
Tc materials and has led to many proposals for unconventional
pairing symmetries. For singlet pairing, the lowest possible
unconventional symmetry without breaking rotational symmetry is
$d_{x^2-y^2} \pm id_{xy}$, and mechanism based on the correlation
effects for such symmetry has been proposed\cite
{Baskaran,Lee,Ogata}. Possible spin-triplet $f$-wave pairing was
also proposed based on ferromagnetic fluctuations\cite{Hu, Kuroki,
ogata1}. Experimentally, however, data reported show contradictory
conclusions\cite {Mazin}, indicating the fragility of the
superconductivity in this system. Furthermore, there are evidences
indicating that two pairing symmetries may be involved in this
system \cite{yang, maska, ogata2}. The issue of whether the pairing
symmetry is conventional or not is still unsettled and needs to be
clarified.

In this paper, we shall focus on the data of the upper critical
fields, measured by the specific heat. The specific-heat technique
probes the bulk properties of the samples and has been proved to be
a powerful tool for investigating the pairing state of many
superconductors\cite{specific}.
Specifically, for the hydrate cobaltate Na$_{0.35}$CoO$_{2} \cdot y$H$_{2}$%
O, upward curvature (a kink structure in the slope) of $H_{c2}$ was
observed \cite{yang, maska}. Similar structure was also observed in
early studies of high Tc materials\cite{highTc}. Based on the
Ginzburg-Landau theory, Joynt\cite {Joynt} attributed the upward
curvature to the transition between two different pairing symmetries
with different critical temperatures. However, in a later
investigation based on microscopic formulation of the gap equation
\cite{Ting}, negative results were found, indicating that the upward
curvature is not due to mixing of two pairing order parameters. The
reason why two approaches give different results lies in the fact
that in the
Ginzburg-Landau theory, $H_{c2} \propto T-T_c$ and phenomenologically, both $%
T_c$ and slopes of $H_{c2}$ are often chosen arbitrarily. If larger
slope of $H_{c2}$ is chosen for smaller $T_c$, $H_{c2}$ of two
pairing symmetries near their $T_c$'s essentially shows the
intersection of two straight lines. A kink in the slope thus arises
and the upward curvature can be easily simulated. In real materials,
however, the slope of $H_{c2}$ and $T_c$ both depend on microscopic
details and are not independent from each other. In fact, in the
Gorkov's microscopic derivation of the Ginzburg-Landau
equation\cite{fetter}, the slope is proportional to $m^*
T_c/ \epsilon_F$ with $m^*$ being the effective mass of the electron and $%
\epsilon_F$ being the Fermi energy. For a single band, $m^*/
\epsilon_F$ are the same for different pairing symmetries, hence
smaller $T_c$ goes with smaller slope, in the opposite trend adopted
in the Ginzburg-Landau equation. Hence in this case, joining two
pairing symmetries with different $T_c$ would not yield the upward
curvature. This picture essentially explains why the upward
curvature is not reproduced in the calculation of Kim et al.
\cite{Ting} for high Tc materials. Then, what is the origin for the
upward curvature in hydrate colaltate ? It is known that
multi-orbitals near the Fermi surface might be involved for the
occurrence of superconductivity\cite{Singh}. Hence $m^*/ \epsilon_F$
can no longer be treated as a fixed parameter for different pairing
symmetries if different pairings occur on different bands. Indeed,
our calculation in below shows that upward curvature feature can
result from the two-band calculation in which different values of
$m^*/ \epsilon_F$ are assumed for different energy bands where
different pairing symmetry occurs. Furthermore, mixing of different
pairing symmetries results in coupling of the ground state to
different Landau levels in the presence of magnetic fields and
causes different upward curvatures.  By direct comparison of
experimental data with $H_{c2}$ obtained by mixing of the lowest
possible pairing symmetries, it is possible to pin down the pairing
symmetries of the system. We find that the upward curvature observed
in the experimental data is
mostly consistent with a transition of the pairing symmetry from $s$ to $%
d_{x^2-y^2}+ id_{xy}$.

We start by first considering pairing that occurs in two bands. The
two-dimensional BCS-like Hamiltonian can be written as
\begin{eqnarray}
&&\hat{H}=\sum_{i=1}^{2}\sum_{k_{i},\sigma }\xi _{k_{i}}\hat{C}_{k_{i}\sigma
}^{+}\hat{C}_{k_{i}\sigma }  \nonumber \\
&+&\sum_{i,j}\sum_{k_{i},k_{j}^{\prime }}V_{k_{i}k_{j}^{\prime }}\hat{C}%
_{k_{i}\uparrow }^{+}\hat{C}_{-k_{i}\downarrow }^{+}\hat{C}_{-k_{j}^{\prime
}\downarrow }\hat{C}_{k_{j}^{\prime }\uparrow },  \label{eq1}
\end{eqnarray}
which was first investigated by Suhl, Matthias, and Walker\cite{Suhl}. Here $%
\mathbf{k}_{1}$ and $\mathbf{k}_{2}$ are wave vectors on two Fermi surface
sheets indexed by 1 and 2. The electron-electron interaction $%
V_{\mathbf{k}_{1}\mathbf{k}_{1}^{\prime }}$ and
$V_{\mathbf{k}_{2}\mathbf{k}_{2}^{\prime }}$ are the intrasheet
contributions and $V_{\mathbf{k}_{1}\mathbf{k}_{2}^{\prime}}$ is an
intersheet contribution. In general, the superconducting pairing
symmetry is different
for different band. Hence we can write the projection of the interaction in the form $%
V_{\mathbf{k}_{1}\mathbf{k}_{1}^{\prime }}=V_{\alpha } \hat{\phi}
_{\alpha } (\mathbf{k}_{1}) \hat{\phi} _{\alpha }^{\ast } (\mathbf{k}_{1}^{\prime })$, $%
V_{\mathbf{k}_{2}\mathbf{k}_{2}^{\prime }}=V_{\beta } \hat{\phi} _{\beta } (\mathbf{k}%
_{2}) \hat{\phi} _{\beta }^{\ast }(\mathbf{k}_{2}^{\prime})$, and
$V_{\mathbf{k}_{1}\mathbf{k}_{2}^{\prime }}=V_{I} \hat{\phi}
_{\alpha }(\mathbf{k}_{1}) \hat{\phi} _{\beta }^{\ast
} (\mathbf{k}%
_{2}^{\prime })$. Here $\alpha $ and $\beta $ are indices for the
pairing symmetry. $\hat{\phi} _{\alpha }(\mathbf{k})$ is an operator
that projects out the corresponding pairing symmetry $\alpha $. For
the lowest possible pairings, we shall consider possible mixing of
$p$ and $f$ waves for triplet pairing and mixing of $s$ and $d$ for
singlet pairing. For the triangle lattice, assuming that the
rotational symmetry is not broken, the appropriate $p$ and
$d$ pairing amplitudes are $p_{x}\pm ip_{y}$ and $d_{x^{2}-y^{2}}\pm id_{xy}$%
. The corresponding projection operators are $\hat{\phi} _{s}(\mathbf{k})=1$, $%
\hat{\phi} _{p\pm ip}(\mathbf{k})=\hat{k}_{x} \pm i\hat{k}_{y}$,
$\hat{\phi} _{d \pm id }(\mathbf{k})= \hat{k}_{x}^{2}-\hat{k}_{y}^2
\pm 2i \hat{k}_{x} \hat{k}_{y}$, and $\hat{\phi}
_{f}(\mathbf{k})=\hat{k}_{x}^{3}-3\hat{k}_{x}\hat{k}_{y}^2$. Note
that there are two possible $f$-waves; however, since they are
related by a rotation of $\pi /6$ and it suffices to consider one of
them (see below for details). Following Kim et al.\cite{Ting}, the
real-space linearized gap equation in the presence of the magnetic
field $\mathbf{\nabla }\times \mathbf{A}$ can be written
as\cite{Ting}
\begin{eqnarray}
&&\hspace{-0.5cm}\Delta _{\alpha }(\mathbf{R})=V_{\alpha
}\sum_{\omega }\int d \mathbf{r} \hat{\phi} _{\alpha }^{\ast }( \phi
) \hat{\phi}
_{\alpha } (\phi)\hat{K}_{1}(\mathbf{r},\omega )\Delta _{\alpha }(\mathbf{R})  \nonumber  \\
&&+V_I \sum_{\omega }\int d\mathbf{r} \hat{\phi} _{\alpha }^{\ast }
(\phi) \hat{\phi} _{\beta } (\phi)\hat{K}_{2}(\mathbf{r},\omega
)\Delta _{\beta }(\mathbf{R}), \label{gap1}
\end{eqnarray}
and \begin{eqnarray} &&\hspace{-0.5cm}\Delta _{\beta
}(\mathbf{R})=V_{\beta }\sum_{\omega }\int d\mathbf{r} \hat{\phi}
_{\beta }^{\ast } (\phi) \hat{\phi} _{\beta } (\phi) \hat{K}_{2}(\mathbf{r},\omega )\Delta _{\beta }(\mathbf{R})  \nonumber  \\
&&+V_I \sum_{\omega } \int d \mathbf{r} \hat{\phi} _{\beta }^{\ast }
(\phi) \hat{\phi} _{\alpha } (\phi) \hat{K}_{1}(\mathbf{r},\omega )
\Delta_{\alpha }(\mathbf{R}). \label{gap2}
\end{eqnarray}
Here $\mathbf{R}$ represents the position for the center of mass of
the Cooper pair, $\mathbf{r}$ is the displacement of the Cooper pair
and $\phi$ is the corresponding angle of $\mathbf{r}$. $\Delta$'s
are pairing amplitudes in real space and $\hat{\phi} _{\alpha
}(\phi)$ are
the projection operators in real space: $\phi _{s}(\phi)=1$, $%
\phi _{p\pm ip}(\phi)= e^{\pm i  \phi}$, $\phi _{d \pm id}(\phi)=
e^{\pm 2i \phi}$, and $\phi _{f}(\phi)=\cos^{3} \phi -3\cos \phi
\sin^2 \phi = cos 3 \phi$. $\hat{K}_{i}(\mathbf{r}, \omega)$ is the
kernel operator, given by
\begin{eqnarray}
\hat{K}_{i}(\mathbf{r}, \omega)= K^0_{i}(\mathbf{r}, \omega)
\exp\left[ \mathbf{r} \cdot (\mathbf{\bigtriangledown}_\mathbf{R}
+\frac{2ie}{\hbar c} \mathbf{A} (\mathbf{R}))\right]
\end{eqnarray}
with
\begin{eqnarray}
K^0_i (\mathbf{r},\omega ) = k_B T N_i (0)^2 \frac{2\pi}{k_{Fi}}
\frac{\exp \left[ \frac{-2r| \omega|} {v_{Fi}} \right]}{r},
\end{eqnarray}
where $k_{Fi}$, $v_{Fi}$ and $N_i(0)$ are the Fermi wavenumber,
Fermi velocity and the two dimensional density of state for the
$i$th band. In the absence of $V_I$, Eqs.(\ref{gap1}) and
(\ref{gap2}) decouples, and their solutions for constant $\Delta$
and $\mathbf{A} =0$  yield relations between $T_c$ and
$V_{\alpha}$\cite{Helfand}
\begin{eqnarray}
\frac{1}{\Gamma_i V_i  N_i}= \sum_\nu \frac{1}{ |2\nu+1|} - \ln
\frac{T^i_{c}}{T}, \label{dos}
\end{eqnarray}
where $i = \alpha$ or $\beta$ with $T^i_{c}$ being the corresponding
transition temperature in zero field  and $\Gamma_i= \int_0^{2\pi}
\frac{d \phi}{2 \pi} |\hat{\phi}_i (\phi)|^2$. Dividing
Eqs.(\ref{gap1}) and (\ref{gap2}) by $V_{\alpha}$ and $V_{\beta}$
respectively and using Eq.(\ref{dos}), $V_{\alpha}$ and $V_{\beta}$
can be eliminated. Following Ref.\cite{Helfand}, if we further adopt
dimensionless variables, $\mathrm{t_i =T/T_c^i}$, $\mathbf{r} =
\mathbf{\rho} /\sqrt{2eH/\hbar c}$ and $h_i =2eH/\hbar c(\hbar
v_{Fi}/2\pi k_BT_c^i )^2 $, the gap equations become
\begin{eqnarray}
& & \Gamma_{\alpha} \left( \sum_\nu\frac{1}{ |2\nu+1|} -
\ln\frac{1}{t_{\alpha}} \right) \Delta_{\alpha}(\mathbf{R}) =
\nonumber \\
& & \hat{K}_{\alpha \alpha} \Delta_{\alpha} ( \mathbf{R}) -
\xi_{\alpha} \gamma \hat{K}_{\alpha \beta} \Delta_{\beta} ( \mathbf{R}) \label{gap3} \\
& &\Gamma_{\beta} \left( \sum_\nu\frac{1}{ |2\nu+1|} -
\ln\frac{1}{t_{\beta}} \right)  \Delta_{\beta}(\mathbf{R}) =
\nonumber \\
& &\hat{K}_{\beta \beta} \Delta_{\beta} ( \mathbf{R}) -
\frac{\xi_{\beta}}{\gamma} \hat{K}_{\beta \alpha } \Delta_{\alpha} (
\mathbf{R}), \label{gap4}
\end{eqnarray}
where $\xi_i = V_I / V_{i}$ and $\gamma = N_{\beta} / N_{\alpha}$.
The operators $\hat{K}$ are given by
\begin{eqnarray}
&& \hat{K}_{nm} = \frac{1}{2\pi}\frac{t_m}{\sqrt{h_m}}\sum_\omega
\int_0^\infty d\rho
\hspace{0.1cm}\exp\left[\frac{-\rho}{\alpha^m_{\omega}}\right]\hspace{0.1cm}\exp
\left[\frac{-\rho^2}{4}\right]\nonumber \\
&& \times \int_0^{2\pi}d\phi \hat{\phi} _{n}^{\ast }(\phi)
\hat{\phi} _{m}
(\phi)\exp\left[\frac{-\rho}{\sqrt{2}}\hat{a}^+e^{i\phi}\right]
\exp\left[\frac{\rho}{\sqrt{2}}\hat{a}^- e^{-i\phi}\right],
\nonumber
\\
\label{K}
\end{eqnarray}
where $\alpha^m_{\omega} = \frac{\sqrt{\frac{2eH}{\hbar
c}}}{2|\omega|}v_{Fm}$ and $\hat{a}^{\pm}= \sqrt{\frac{\hbar
c}{4eH}} [(\bigtriangledown+2ieA)_x \pm i( \bigtriangledown+2ieA)_y
]$.

Eqs.(\ref{gap3}) and (\ref{gap4}) are the governing equations for
the situation when two pairing symmetries occur on different energy
bands. However, it also covers the case when the two pairing
symmetries occur in the same energy band. In that case, one simply
sets $\gamma =1$, $\xi_i =1$, and drop the index $i$.
Eqs.(\ref{gap3}) and (\ref{gap4}) can be solved by expanding
$\Delta_{\alpha}$ and $\Delta_{\beta}$ in terms of the set of
eigenfunctions $| \psi_n \rangle$ of the two-dimensional Schrodinger
equation in the presence of $\mathbf{A} =(0,Hx,0)$
\begin{eqnarray}
\Delta_i = \sum_{n=0}^{\infty} A^i_n | \psi_n \rangle,
\end{eqnarray}
where $i = \alpha$ or $i=\beta$. Note that the eigenfunctions $|
\psi_n \rangle$ are essentially the Landau levels. The operators
$\hat{K}$ couple different Landau levels. The detailed coupling is
determined by the bi-product projection operator, $\hat{\phi}
_{n}^{\ast } (\phi) \hat{\phi} _{m} (\phi)$, in Eq.(\ref{K}), which,
after being integrated, selects the correct combinations of
$\hat{a}^+$ and $\hat{a}^-$ that survive. The selected combination
of $\hat{a}^+$ and $\hat{a}^-$ then determines how $A^i_n$ couple.
For the mixing of $s$-wave and $d+id$, $\{ A^s_n \}$ couples with
$\{ A^{d+id}_{n+2} \}$; while for the mixing of $s$-wave and $d-id$,
$\{ A^s_n \}$ couples with $\{ A^{d+id}_{n-2} \}$. On the other
hand, for the mixing of $p \pm ip$ and $f$-wave, because $\hat{\phi}
_{f}(\phi)$ contains both $e^{3i \phi}$ and $e^{-3i \phi}$, in
addition to the coupling between $A^{p \pm ip}_n$ and $A^f_n$, there
are also couplings among $\{ A^f_n \}$. Since the bi-product
projection operator in $\hat{K}_{ff}$, $\hat{\phi} _{f}^{\ast }
(\phi) \hat{\phi} _{f}(\phi)$, contains $e^{\pm i 6 \phi}$, $A^f_n$
couples with $A^f_{n+6}$ for each $n$; while because the bi-product
projection operators in off-diagonal $\hat{K}_{\alpha \beta}$
($\alpha \neq \beta$), contains $e^{\pm i4 \phi}$ and $e^{\pm i 2
\phi}$, we found that $\{ A^f_n \}$ couple with $\{ A^{p+ip}_{n+4}
\}$ and $\{ A^f_n \}$ couple with $\{ A^{p-ip}_{n+2} \}$. The
coupling coefficients are most conveniently expressed in terms of
the following functions\cite{Ting}
\begin{equation}
\beta^{i}_n=\frac{t_{i}}{\sqrt{h_{i}}}\int_0^\infty d\rho\hspace{0.2cm}\frac{%
e^{-\frac{\rho^2}{4}}L_n\left(\frac{\rho^2}{2}\right)-1}{\sinh\left(\frac{%
t_{i}\rho}{\sqrt{h_{i}}}\right)}
\end{equation}
and
\begin{equation}
\alpha^{i}(2k,n)=\frac{t_{i}}{\sqrt{h_{i}}}\int_0^\infty d\rho\hspace{0.2cm}%
\frac{e^{\frac{-\rho^2}{4}}\left(\frac{-\rho^2}{2}\right)^kL^{2k}_n\left(%
\frac{\rho^2}{2}\right)}{\sinh\left(\frac{t_{i}
\rho}{\sqrt{h_{i}}}\right)},
\end{equation}
where $i$ is the index for the pairing symmetry, $L_n(x)$ is the
Laguerre polynomial and $L^{2k}_n$ is the associated Laguerre
polynomial in Rodrigues representation. For triplet pairing, we find
that recursion relations for the mixing of $f$-wave with $p_x \pm
ip_y$ are given by
\begin{eqnarray}\label{coef1}
&&\nonumber
\frac{-1}{2}\sqrt{\frac{(n-6)!}{n!}}\alpha^f(6,n-6)A^{f}_{n-6}+
\left(\ln\frac{1}{t_f}+\beta^f_n\right)A^{f}_n\\
&&\nonumber-\frac{1}{2}\sqrt{\frac{(n+6)!}{n!}}\alpha^f(-6,n+6)A^{f}_{n+6}
\\
&&+
\xi_f \gamma\sqrt{\frac{(n \mp 2)!}{n!}}\alpha^{p \pm ip}( \pm 2,n \mp 2)A^{p \pm ip}_{n \mp 2} \nonumber \\
&&-\xi_f \gamma\sqrt{\frac{(n \pm 4)!}{n!}}\alpha^{p \pm ip} ( \mp
4,n \pm 4)A^{p \pm ip}_{n \pm 4}=0, \label{mix1}
\end{eqnarray}
and
\begin{eqnarray}\label{coef2}
&&\hspace{-1cm}\nonumber\left(\ln\frac{1}{t_p}+
\beta_n^p\right)A^{p \pm  ip }_n-\frac{\xi_{p \pm ip} }{2\gamma}\sqrt{\frac{(n \mp 4)!}{n!}}\alpha^f(\pm 4,n \mp 4)A^{1f}_{n \mp 4}\\
&&+\frac{\xi_{p \pm ip} }{2\gamma}\sqrt{\frac{(n \pm
2)!}{n!}}\alpha^f( \mp 2,n \pm 2)A^{f}_{n \pm 2}=0. \label{mix2}
\end{eqnarray}
On the other hand, for singlet pairing, recursion relations for the
mixing of $s$-wave with $d_{x^2-y^2} \pm id_{xy}$ are given by
\begin{eqnarray}\label{coef5}
&&\hspace{-1cm}
\left(\ln\frac{1}{t_s}+\beta^s_n\right)A^{s}_n+\frac{\xi_s}{\gamma}\sqrt{\frac{(n
\pm 2)!}{n!}}\alpha^{d \pm id} ( \mp 2,n \pm 2)A^{d \pm id }_{n \pm
2}=0, \nonumber \\ \label{mix3}
\end{eqnarray}
and
\begin{eqnarray}\label{coef6}
&&\hspace{-1cm} \left(\ln\frac{1}{t_d}+\beta^d_n\right)A^{d \pm id
}_n+\xi_{d \pm id} \gamma\sqrt{\frac{(N \mp 2)!}{n!}}\alpha^s(\pm
2,n \mp 2)A^{s}_{n \mp 2}=0. \nonumber \\ && \label{mix4}
\end{eqnarray}

When applying the above analysis to the calculation of $H_{c2}$ with
mixing of two pairing symmetries, $\alpha$ and $\beta$, one assumes
that the pairing symmetries are associtaed with different $T_c$'s
with $T^{\alpha}_c
> T^{\beta}_c$. In other word, the pairing symmetry $\alpha$ is the
stable bulk pairing state at low fields while $\beta$ is the stable
pairing state at high fields. Therefore, one starts from
$\Delta_{\alpha} = A^{\alpha}_0 | \psi_0 \rangle$. Mixing to the
other symmetry, $\beta$, then couples $A^{\alpha}_0$ to
$A^{\beta}_n$ with $n \geq 1$. Since $\{ A^s_n \}$ couples with $\{
A^{d \pm id}_{n \pm 2} \}$, this analysis implies that one can only
have the transitions from high-temperature $s$-wave to
low-temperature $d+id$ or from high-temperature $d-id$ to
low-temperature $s$-wave. On the other hand, for the mixing of $p
\pm ip$ and $f$-wave, because $\hat{\phi} _{f}(\phi)$ contains both
$e^{3i \phi}$ and $e^{-3i \phi}$, both transitions from
high-temperature $p \pm ip$ to low-temperature $f$-wave and from
high-temperature $f$-wave to low-temperature $p \pm ip$ are
possible. Note that there are two possible $f$-waves,
$\hat{k}_{x}^{3}-3\hat{k}_{x}\hat{k}_{y}^2$ and
$\hat{k}_{y}^{3}-3\hat{k}_{y}\hat{k}_{x}^2$. Since the two $f$-waves
are related by exchanging $\hat{k}_{x}$ and $\hat{k}_{y}$ which
simply exchanges $p_x + ip_y$ and $p_x - i p_y$, it suffices to
consider one of them.

When solving $H_{c2}$, it is important to note that there are many
eigenvalues $H(T)$ satisfying Eqs.(\ref{gap3}) and (\ref{gap4}) and
only the largest one defines $H_{c2}$. To compare the calculated
$H_{c2}$ with experimental data, one needs to fix scales of
temperatures and magnetic fields. The transition temperature $T_c$
of the most stable bulk pairing state determines the temperature
scale. On the other hand, the scale of magnetic fields can be fixed
by the data points with lower magnetic fields. The remaining
parameters are the ratio of Fermi velocities, the ratio of $T_c$,
$\xi_i$ and $\gamma$.  At this point, it is important to note that
according to Eqs. (\ref{mix1})-(\ref{mix4}), for different mixing
scheme, different Landau levels are mixed. As a result, different
mixing scheme results in different upward curvature. To find the
best fit to the experimental data, after fixing scales of
temperatures and magnetic fields, we vary the remaining parameters
to find the best. We find that the transition from $s$-wave at low
fields to $d+id$ at high fields is the best fit to the data. In Fig.
1, we show numerical results of $H_{c2}$ for the transition from
$s$-wave at low fields to $d+id$ at high fields in comparison with
experimental data obtained by specific-heat measurement. The fitting
parameters are the ratio of Fermi velocities $v^s_F/v^{d+id}_F =
0.8$, $T^{d+id}_c/ T^s_c = 0.5$, $\xi_s=4.3$, $\xi_{d+id}=0.89$, and
$\gamma=1.1$. These values are in reasonable regime. The close
fitting ro the experimental data clearly shows that singlet pairing
dominates in hydrate cobaltat, which is also consistent with recent
NMR data\cite{nmr}. Furthermore, it implies that two energy bands
are involved and supports results based on LDA
calculations\cite{Singh} where two bands constructed from the three
Co $t_{2d}$ orbitals intersect the Fermi level.
\begin{figure}[ht]
\rotatebox{-90}{\includegraphics[width=60mm]{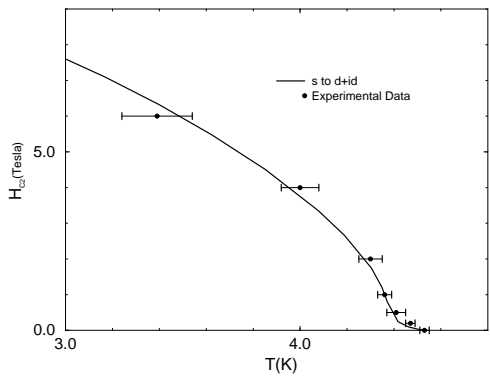}}\caption{\small
The comparison of experimental data of $H_{c2}$\cite{yang} with
numerical results based on the transition from $s$-wave at low
fields to $d+id$ at high fields.}
\end{figure}

In conclusion, we have investigated the origin for the upward
curvature of the upper critical field ($H_{c2}$)
observed in hydrate cobaltate Na$_{0.35}$CoO$%
_{2} \cdot y$H$_{2}$O. Analysis based on the microscopic gap
equation shows that the observed upward curvature results from the
transition between two different pairing symmetries that occur on
different energy bands. Furthermore, it is found that the transition
of the pairing symmetry from $s$-wave at low fields to $d_{x^2-y^2}+
id_{xy}$ at high fields is the best fit to the experimental data.

This work is supported by the NSC of Taiwan. We thank Prof. Hsiu-Hau
Lin for useful discussions.


\end{document}